\documentclass[sigconf]{acmart}
\definecolor{highlight}{RGB}{0, 0, 200}
\usepackage{algorithm}
\usepackage{algpseudocode}
\usepackage{makecell}
\usepackage{multirow}
\usepackage{caption}
\usepackage{subcaption}

\usepackage{titlesec}
\titlespacing*{\section}{0pt}{4pt plus 1pt}{4pt minus 1pt}
\titlespacing*{\subsection}{0pt}{4pt plus 1pt}{4pt minus 1pt}

\AtBeginDocument{%
  \providecommand\BibTeX{{%
    \normalfont B\kern-0.5em{\scshape i\kern-0.25em b}\kern-0.8em\TeX}}}

\begin{document}
\title{Reward-free Policy Imitation Learning for Conversational Search}

\author{Zhenduo Wang}
\email{zhenduow@cs.utah.edu}
\orcid{1234-5678-9012}
\affiliation{%
  \institution{University of Utah}
  %\city{Salt Lake City}
  %\state{Utah}
  %\country{United States}
  %\postcode{84112}
}

\author{Zhichao Xu}
\email{zhichao.xu@utah.edu}
\affiliation{%
  \institution{University of Utah}
  %\city{Beijing}
  %\state{Utah}
  %\country{China}
  }

\author{Qingyao Ai}
\email{aiqy@tsinghua.edu.cn}
\affiliation{%
  \institution{Tsinghua University}
  %\city{Beijing}
  %\state{Utah}
  %\country{China}
  }


\begin{abstract}

Existing conversational search studies mainly focused on asking better clarifying questions and/or improving search result quality. These works aim at retrieving better responses according to the search context, and their performances are evaluated on either single-turn tasks or multi-turn tasks under naive conversation policy settings. This leaves some questions about their applicability in real-world multi-turn conversations where realistically, each and every action needs to be made by the system itself, and search session efficiency is often an important concern of conversational search systems. While some recent works have identified the need for improving search efficiency in conversational search, they mostly require extensive data annotations and use hand-crafted rewards or heuristics to train systems that can achieve reasonable performance in a restricted number of turns, which has limited generalizability in practice. 

In this paper, we propose a reward-free conversation policy imitation learning framework, which can train a conversation policy without annotated conversation data or manually designed rewards. The trained conversation policy can be used to guide the conversational retrieval models to balance conversational search quality and efficiency. To evaluate the proposed conversational search system, we propose a new multi-turn-multi-response conversational evaluation metric named Expected Conversational Reciprocal Rank (ECRR). ECRR is designed to evaluate entire multi-turn conversational search sessions towards comprehensively evaluating both search result quality and search efficiency.
\end{abstract}



\settopmatter{printfolios=true}
\maketitle
\section{Introduction}

Conversational AI is one of the long-standing research goals in both natural language processing (NLP) and information retrieval (IR) communities as well as industry. Along with the development of conversational agents such as Microsoft Cortana, conversational search and other related research topics such as conversational question answering, task-oriented dialog systems, social chat, and not the least conversational recommendation have drawn many interests in the past few years \cite{hauff2021conversational, balog2021conversationalchallenge, zhang2020recent, zaib2021conversationalqasurvey}. Among these topics, conversational search is a retrieval-based approach for conversational information seeking. It retrieves responses given the initial query and current conversation context \cite{zamani2022conversational}. Most existing conversational search studies and datasets \cite{InforSeek_Response_Ranking, udcdataset, yang2020iart, penha2019mantis, zhang2018modeling, henderson2019repository} cast conversational search as a next-response ranking (NPR) task where all types of responses (e.g., returning results, and asking clarifying questions) are indiscriminately sampled and ranked together as response candidates. 

Recent studies about conversational search conceptualization \cite{radlinski201framework, azzopardi2018conceptualizing}, utterance intent identification \cite{InforSeek_User_Intent, InforSeek_User_Intent_Pred}, and asking clarifying questions \cite{aliannejadi2019asking, kiesel2018clarifyingsatisfaction, bi2021asking} have suggested that modeling conversational search as naive sequence of responses as in NPR maybe not be the most effective way. Rather, it would be better to categorize user and system utterances separately using fine-grained conversation action spaces. For example, user can perform actions such as inquiry, refinement, and closing, while system can perform actions such as revealing and clarification. This naturally requires the system to have the ability to decide which of these actions to perform (conversational search policies), which leads to a system architecture that is very similar to a task-oriented dialog system \cite{gao2022neural}, and MACAW \cite{zamani2020macaw} is a virtual example of such system architecture. However, within the scope of conversational search, only a few works have been proposed in this direction, or at least have addressed the need of making conversation action decisions, such as \cite{zhang2018towards, aliannejadi2020convai3, aliannejadi2021building}. Wang and Ai \cite{wang2021controlling, wang2022simulating} studied the risk of conversational search and proposed to solve conversational search policy as a risk-controlling task. However, the risk of asking clarifying questions is not the only factor of conversational search policy. Other factors such as how to maximize search efficiency and improve user engagement during conversation should also be considered. Hence, their conversational search policy could be risk-biased and overlook the opportunities to improve search result quality.

There are more studies about dialog policy learning in the task-oriented dialog systems community. Many existing works about policy learning employ reinforcement learning (RL) algorithms and train a dialog policy model through interactions with user simulator models \cite{fatemi2016policy, dhingra2016towards, peng2017composite, su2018discriminative}. However, designing reinforcement learning rewards is a big challenge in these works since it requires a large amount of knowledge about the task and dataset. Poorly designed rewards can often lead to undesired system behaviors \cite{takanobu2019gdpl}. In order to achieve good performances, most of these works manually tune their rewards, which consequently limits their generalization ability. 

In this paper, we propose a novel reward-free conversational search policy imitation learning (IL) framework. Our IL framework trains a simple conversation policy model, which decides between asking the retrieved clarifying question and returning the retrieved results to the user at each conversation stage, given the current conversation context and retrieval results. Inspired by recent advances in Generative Adversarial Imitation Learning (GAIL) \cite{ho2016generative}, to train the conversation policy model, we compute and identify an expert trajectory (a multi-turn-multi-response tuple) from all possible conversation trajectories using an automatic evaluation metric, and we use the expert trajectories as the training samples without any data annotations or manually-crafted rewards. Our IL framework also allows us to train policies according to different conversational search evaluation metrics that represent different user assumptions. Hence, our policy learning framework could potentially generalize to other future conversational search tasks, datasets, and user assumptions.

To better evaluate the multi-turn-multi-response conversational search sessions, we also propose a new automatic conversational evaluation metric called Expected Conversational Reciprocal Rank (ECRR). Our metric can evaluate the entire response trajectory of the search agent in the multi-turn conversational search session. This is fundamentally different from naive metrics which evaluate one single-turn retrieved response quality at a time. Our metric is also an attempt toward a comprehensive evaluation metric for conversational search, which evaluates both the search result quality and search session efficiency.

We consider our work has three contributions: 
\begin{itemize}
\vspace{-3pt}
    \item We propose a new reward-free conversational search policy imitation learning framework, which can learn a conversation policy to jointly optimize search effectiveness and efficiency without data annotations or manually designed rewards. %The learned conversation policy can be used to guide the conversational retrieval models to maximize both conversational search effectiveness and efficiency.
    \item To comprehensively evaluate search effectiveness and efficiency, we propose a new multi-turn-multi-response conversational search evaluation metric named Expected Conversation Reciprocal Rank, which can better approximate system performances in real-world scenarios. 
    \item To the best of our knowledge, we are also the first to use multi-turn metrics directly as an indirect learning goal of conversational policy learning.
\vspace{-3pt}
\end{itemize}

The rest of the paper is organized as follows:
we first review the related work (\S\ref{sec:related}); then introduce the proposed IL-based framework (\S\ref{sec:algorithm}) and propose to evaluate conversational search task with the proposed ECRR metric (\S\ref{sec:metric}). 
We cover the detailed experimental setup (\S\ref{sec:experiment}), report and analyze the experimental results (\S\ref{sec:result}).
Finally, we wrap this work up and point out future directions (\S\ref{sec:conclusion}).
To facilitate reproducibility of the IR community, our code will be made public once this work is accepted.

% We will detailedly explain our contribution 1 and 3 in \S\ref{sec:algorithm}, contribution 2 in \S\ref{sec:metric}. 

\section{Related Works}
\label{sec:related}
\subsubsection*{\textbf{Conversational Search}}

Studies on conversational search can be traced back to the very beginning of research on interactive information retrieval \cite{belkin1995earlyroots, croft1987i3r,oddy1977earlyroots2}. The framework proposed by Radlinski and Craswell \cite{radlinski201framework}, along with other conceptualization works including \cite{azzopardi2018conceptualizing, deldjoo2021multimodal} is some examples of the recent surge of conversational search system studies. These works have taken various approaches and covered multiple aspects of conversational search. For example, Yu et al. \cite{yu2020queryrewriting, yu2021few} study effective query rewriting and learning contextualized query representation from an ad hoc teacher model. Zhang et al. \cite{zhang2018towards} study conversational preference elicitation. Vakulenko et al. \cite{vakulenko2019knowledge} study knowledge-based conversational search. Some works \cite{aliannejadi2019asking, choi2018quac, zamani2020mimics, InforSeek_Response_Ranking} provide datasets with different focuses for these studies. Other works such as \cite{anand2020seminar,hauff2021conversational, fu2020tutorial, gao2018neural} provide seminars and tutorials from different standpoints.

Among these topics, a popular line of work (e.g. \cite{rao2019answer, zamani2020analyzing, aliannejadi2019asking}) about asking clarifying questions is highly related to this paper. %(This part could be extended and introduced with more details when we feel it is needed) 
Asking clarifying questions in conversational search can be traced back to the TREC 2004 HARD track \cite{trechard}, where asking clarifying questions is a system option to get additional information. Recently, more works have shown that asking clarifying questions can benefit conversational search systems. A study in 2018 \cite{kiesel2018clarifyingsatisfaction} showed that users enjoyed interacting with conversational systems. Following this work, several other works \cite{aliannejadi2019asking, zamani2020generating, penha2020challenges} demonstrate that asking clarifying questions is a convenient alternative when the query information is insufficient for retrieving good results. We may need to reconsider the question today that how long we can be optimistic about users' tolerance and patience for the mistakes of our conversational systems after their freshness for conversational AI, as interacting with conversational AI will soon if not has already become a norm of human life. %(citation?) 
However, only a few works \cite{aliannejadi2020convai3, xu2019asking, wang2021controlling, wang2022simulating} have addressed the problem of deciding whether to ask clarifying questions in conversational search systems. Our work extends these works and generalizes the problem of identifying the need of asking clarifying questions to conversational search policy modeling.

\subsubsection*{\textbf{Conversational Search Evaluation}}
The evaluation of conversational search remains open and diverse because of (1) conversational intelligent system being a common research interest for both natural language processing and information retrieval community, (2) the term-sharing between conversational search and other related areas like dialog system and conversational QA, (3) the variety of task configurations, solutions, and evaluations (e.g. slot filing/response retrieval/generation, single/multiple responses per turn, single/multi-turn evaluation), and more. Due to the cost and complexity of involving human users in the loop, apart from a few works which study the online evaluation framework and interface \cite{kelly2009questionnaires, kaushik2021conceptual, azzopardi2018conceptualizing}, most works simulate users' behaviors and evaluate the conversation sessions with offline methods for its simplicity and reproducibility. Metrics evaluated on single-turn single-response can be categorized as word-overlap-based (BLEU \cite{papineni2002bleu},  METEOR \cite{banerjee2005meteor}), embedding-based (BERT-Score \cite{zhang2019bertscore}), learning-based (BERT-RUBER \cite{ghazarian2019bertruber}), or F1 score for slot filing tasks. Metrics evaluated on single-turn multi-response are mainly ranking-based metrics (nDCG \cite{jarvelin2002sdcg}, RBP \cite{moffat2008rbp}, ERR \cite{chapelle2009err}).

The evaluation of conversational search still faces many challenges \cite{penha2020challenges}. Metrics evaluated on multi-turn sessions are usually done by combining single-turn metrics with or without session-based weighting (SWF \cite{liu2018swf}, sDCG \cite{jarvelin2002sdcg}). As a matter of fact, the adaption of single-turn evaluation metrics for multi-turn evaluation is actually more complicated than it shows in the above methods. The reason is that a conversational search system which, by its definition, will interact with the user and generate indefinite search sessions, which means that different system responses can lead to completely different conversation trajectories. Thus combining single-turn evaluation metrics through weighting seems questionable, as it overlooks the effect of each turn on future turns due to the turn-independence assumption. Recently, there have been some works about multi-turn evaluation. sRBP \cite{lipani2019srbp} evaluates session rank based on user model. Wang and Ai \cite{wang2021controlling, wang2022simulating} evaluate conversational search sessions by simulating with various types of users. ECS \cite{lipani2021howamidoing} suggests to evaluate entire conversation session by identifying sub-topic. We follow the same setup with these works and use a cascade user model for multi-turn evaluation.

While most of the above metrics are focused on evaluating result quality or search effectiveness, other works also mention measuring search efficiency. Optimizing search efficiency (or search effort) means promoting more efficient conversational systems which can achieve the same goal with less interaction with the user, and are usually measured by the number of conversation turns. Early works like \cite{jurafskyspeech, harabagiu2005iqaefficiency} highlight efficiency as another major measure beside effectiveness in interactive QA evaluation. Among more recent works, RBP \cite{moffat2008rbp} and ERR \cite{chapelle2009err} are two single-turn multi-response evaluation metrics with efficiency measuring aspect. They are both based on an explicit user behavior simulator that models the user's patience during the search in exhausting the search result list.

\subsubsection*{\textbf{Task-oriented Dialog System}}
With the ability to interact with the user, conversational search systems are closely related to dialog systems, as they both aim to help the user through multi-turn conversations. However, a task-oriented dialog system usually knows the exact task it needs to solve, such as flight booking or weather query, while conversational search tasks are more open. One way to connect the two tasks and think about their similarities is to regard conversational search as a goal-oriented dialogue, where its goal is information seeking \cite{gao2022neural}. Anand et al. also \cite{anand2020seminar} type conversational search system as \textit{an information-seeking dialogue system with information retrieval capabilities}. Works like these have shown that conversational search and task-oriented dialog systems are highly similar in many aspects. Recently, a system-ask-user-respond scheme that is common in task-oriented dialog systems is also seen in conversational search and recommendation system studies like \cite{zhang2018towards, aliannejadi2019asking}. While task-oriented dialog systems usually respond to the user by generating natural language responses, conversational search systems focus on retrieving relevant information from massive web sources to the user. Besides this difference, a task-oriented dialog system can often involve solving multiple sub-questionnaires of the task, while the conversational search problem we focus on in this paper is to clarify and answer the user's initial information request through multi-turn conversation.

Because of the connections between task-oriented dialog systems and conversational search systems, these systems share many structural similarities. Previous research on task-oriented dialogue systems in NLP can be roughly categorized into two groups \citep{gao2020robust, zhang2020recent}: (1) pipeline/modular system, and (2) end-to-end system. Pipeline systems regard the process of a task-oriented dialogue system as an iterative decision-making process with four major modules: natural language understanding (NLU), dialogue state tracking (DST), dialogue policy (POL), and natural language generation (NLG). During the iterative decision-making process, the system reads the dialogue (NLU) and processes the information to have an understanding of the current dialogue state (DST), then decides its next action concerning the understanding (POL), and finally generates natural language output implementing the action (NLG). Some works try to combine some of the modules such as 
combining NLU with DST \cite{neuraldst1,neuraldst2}, or policy with NLG \cite{policyNLG1,policyNLG2}. 

End-to-end systems such as \cite{end2end1, end2end2, end2end3} model task-oriented dialogue as a response generation problem through language modeling, with the entire dialogue history modeled as a long text sequence, essentially serializing the pipeline.

\subsubsection*{\textbf{Imitation Learning}}
Imitation learning (IL) aims to train a system to mimic human behavior by learning from human demonstrations \cite{hussein2017imitation}. IL can be implemented as supervised learning, where the agent learns to behave like the expert on all training samples. This is known as the behavior cloning method \cite{behaviorcloning}. However, simply copying the behavior on the training set has been shown to be inefficient and not generalizable. IL is more often cast as an inverse reinforcement learning (IRL) problem \cite{abbeel2004apprenticeship, ziebart2008maximum, wulfmeier2015maximumdeep, wulfmeier2015deep}, where the system tries to infer the reasoning/logic behind these expert demonstrations, then learn to apply them to unseen scenarios to generalize. In general reinforcement learning (RL) algorithms, it is usually required to have explicit or manually designed rewards signals for training. However, in real-world tasks such as automatic driving and robotics, such signal rarely exists and defining rewards is usually difficult or even impossible. IRL is a learning paradigm to learn a policy with no available rewards by inferring the rewards from expert demonstrations for the RL algorithm to use. 

The IRL algorithms and their deep neural network versions iteratively perform the IRL-RL learning steps from expert demonstrations and are generally considered to be costly in terms of time. With the advance of studies about generative adversarial nets (GAN) \cite{goodfellow2014generative}, Finn et al. \cite{finn2016connection} suggest the equivalencies between GAN, IRL, and energy-based models, and Ho et al. \cite{ho2016generative} propose that the IRL-RL learning iteration can be simplified by solving the dual problem of occupancy measure matching. They subsequently propose a generative adversarial imitation learning (GAIL) algorithm, which could reduce the cost of imitation learning by a large margin. Till today, GAIL has been studied and extended by many works \cite{baram2017endtoendgail,song2018multigail, torabi2018gailfromobservation}. 

The applications of imitation learning methods can mostly be found in robotics and autonomous driving. To the best of our knowledge, our work is among the very first efforts to apply imitation learning methods to conversational search system studies.

\section{Proposed System}
\label{sec:algorithm}
We start this section with an overview of our conversational search system (\S\ref{sec:overview}); then we walk through the system from retrieval models (\S\ref{sec:retrieval}) to the policy model (\S\ref{sec:policy}). Finally we introduce the inference of the proposed model (\S\ref{sec:inference}).
% In this section, we introduce our conversational search system with a policy model and the imitation learning algorithm we use to train the policy model.

\subsection{Conversational Search System with Policy}
\label{sec:overview}
Our conversational search system models a conversational search agent which aims to retrieve relevant information for the user's information need by interacting with the user through multi-turn conversation. Our system consists of two modules:

The first module is a collection of retrieval models, each modeling a specific action that the agent may take, e.g. retrieving results, or asking clarifying questions. We assume that most conversational search systems such as \cite{xu2019asking, aliannejadi2020convai3} have the same or similar structure as the first module. However, most of them do not have a model to decide which action to make, which is our second module. 

The second module is a conversational search policy model, which decides the next agent action given the user's search query, current conversation context, and the retrieved candidates of each action from the first module. In our work, we only consider two main agent actions, returning the retrieved result to the user or asking a clarifying question to the user. Therefore, our policy model only needs to make decisions between the two actions, and there will be only two retrieval models, namely the result retriever and the clarifying question retriever. However, given enough community attention and studies on other possible agent actions and related datasets, our agent should easily be extended and adapted. Our conversational search agent is also an implementation of recent conversational search conceptualization studies, which suggest categorizing agent actions and modeling them individually as discussed in Section 1.

It has been proposed and shown in \cite{zamani2020macaw, wang2021controlling, wang2022simulating} that integrating the retriever outputs in the policy model can improve the overall agent performance and control the risk in clarifying questions compared to the popular baseline of first predicting agent action and then running the corresponding retrieval model. Following their works, our system also runs the retrieval models before the policy model. Hence, we will first introduce the retrieval models (\S\ref{sec:retrieval}) and then the policy model (\S\ref{sec:policy}).

\subsection{Retrieval Models}
\label{sec:retrieval}
We have two retrieval models for modeling the search result retrieval and clarifying question retrieval tasks, respectively. The goal of the retrieval models is to retrieve the best result or clarifying question-based on the search query and conversation context. The result and clarifying question retriever use the same poly-encoder \cite{polyencoder} structure but do not share parameters and are trained separately.

\subsubsection*{\textbf{Poly-encoder}} Assume the conversation context between user and agent can be represented as $\{U_0,A_0,..,U_N,A_N\}$, where $U$s are user utterances, and $A$s are agent utterances. The response candidates can be represented as $P =\{p_1, .., p_K\}$. In poly-encoder, the appended search context $q= \texttt{[S]} U_0 \texttt{[SEP]} A_0 \texttt{[SEP]} U_1...$ is first encoded into vectors $(h_{q_1},..,h_{q_{2N}})$ with pretrained transformer, where \texttt{[S]} represents the start of sentence token, and \texttt{[SEP]} represents the sentence segmentation token. Then it attends to $m$ learnable codes $(C^1, .., C^m)$ to generate $m$ attended context vectors $(q^1,..,q^m)$:
\begin{equation}
q^i = \sum_j^{2N} w_{j}^i h_{q_j} 
\end{equation}
where $(w_1^{i},.., w_{2N}^{i}) = \text{softmax}(C^i \cdot h_{q_1}, .., C^i \cdot h_{q_{2N}})$.

The poly-encoder then encodes each candidate $p_k$ into a vector $E_{p_k}$ using pretrained transformer $T$ and dimension reduction function $\textit{red}(\cdot)$, which can aggregate a sequence of vectors into one vector:

\begin{equation}
E_{p_k} = \textit{red}(T(p_k))
\end{equation}

Finally, the $m$ context vectors $(q^1,..,q^m)$ attend to the candidate vector $E_{p_k}$ to compute candidate-attended context vector $E_{q}$:
\begin{equation}
E_{q} = \sum_i^m w_iq^i
\end{equation}
where $(w_1,..., w_m) = \text{softmax}(E_{p_k}\cdot q^1, .. , E_{p_k}\cdot q^m)$. Then the ranking score of candidate $p_k$ is computed as the dot product 
$s_k = E_{p_k}\cdot E_{q}$. The output of the retriever is the rank of all the candidates and their ranking scores.

The result retriever and clarifying question retriever are trained separately. Both of them are first initialized using pretrained checkpoints, and then fine-tuned on batches of \texttt{(search context, result)} or \texttt{(search context, clarifying question)} pairs to minimize the cross-entropy loss between the softmax result of ranking score vector and the true relevance label vector.
\begin{equation}
L_{\text{poly}} = \sum^B \text{CE}(\text{softmax}(s_1,..,s_K), I(p_1,..,p_K))
\end{equation}

After the retrieval models are trained, their parameters will be fixed during the training of policy models.

\subsection{Policy Model and Imitation Learning}
\label{sec:policy}
\subsubsection*{\textbf{Markov Decision Process}}
We consider the task of choosing which retrieval result to return in each stage of conversational search as a Markov Decision Process (MDP) problem. The conversational search MDP is a tuple $\{\mathcal{S}, \mathcal{A}, \mathcal{T}, \mathcal{R}\}$ where:
\begin{itemize}
    \item $\mathcal{S}$ is a set of conversational states
    \item $\mathcal{A}$ is a set of system actions
    \item $\mathcal{T}(s, a)$ is a transition probability distribution that an action $a$ in state $s$ will lead to $s'$
    \item $\mathcal{R}(s, a)$ is the immediate reward of taking action $a$ in state $s$
\end{itemize}

To be more specific, the state $\mathcal{S}$ comprises the initial query, utterance history context, and retrieved clarifying questions and results. The system action set $\mathcal{A} = \{\text{return results}, \text{ask question}\}$. The transition probability $\mathcal{T}$ is modeled by a user cascade model, which will be explained in section~\ref{sec:experiment}

A trajectory $\tau$ in MDP is a series of system actions together with their resulting states. $\tau$ can be efficiently represented as $\tau = \{ (s_1, a_1), .., (s_t, a_t) \}$. In conversational search, the system only has two actions. We assume that the action of returning results will end the conversation, and the action of asking a question will continue the conversation. Therefore, any conversational search trajectory will be an arbitrary number of asking-question actions followed by one returning-result action.

\subsubsection*{\textbf{Policy Model}}
The job of our conversational search policy model is to make decisions of which agent action to take, at any state $S$. In our setting, the agent only has two actions, either retrieving result for user's query or asking a clarifying question. Hence, the output of our policy model is a probability distribution over the two actions. To represent the input state $S$, we first use pretrained transformer to encode the initial query, $iq$, the current conversation utterances $q= \texttt{[CLS]} U_0 \texttt{[SEP]} A_0 \texttt{[SEP]} U_1...$, the retrieved result candidates $(\textit{re}_1,..,\textit{re}_K)$, and the retrieved clarifying question candidates $(\textit{cq}_1,..,\textit{cq}_K)$. Then, we concatenate all these encodings together with the retrieved result and clarifying question scores as the state representation $S=(\textit{iq}, q, \textit{re}_1,..,\textit{re}_K, \textit{cq}_1,.., \textit{cq}_K, s_{\textit{re}}^{1:K}, s_{\textit{cq}}^{1:K})$. The policy model $G$ is a 2-layer feed-forward neural net:

\begin{equation}
G(S) = \text{softmax}(W_{G2}\cdot \phi(W_{G1}\cdot S + b_{G1}) + b_{G2})
\end{equation}

where $W_{G1}, W_{G2}, b_{G1}, b_{G2}$ are learnable parameters.

\subsubsection*{\textbf{Learning}}
Policy models are usually trained by reinforcement learning methods such as policy gradient or Q-learning, which require manually designing and tuning the rewards function of each action. For example, Wang and Ai \cite{wang2022simulating} empirically determine the reward of asking clarifying questions and returning answers with the intention to control the risks of agent actions. However, this could lead to a rather conservative conversational search policy, which will further weaken the possibility of improving result quality by asking more clarifying questions.

RL algorithms are notoriously hard to train due to their requirement of manually designing and tuning rewards functions. To avoid this effort, we propose to use imitation learning (IL), which is to learn the policy directly from the expert demonstrations of the task without the help of additional training signals. The imitation learning algorithm we use is Generative Adversarial Imitation Learning (GAIL), which alternatively updates and optimizes a (state, action) evaluator as a discriminator $D$ and the policy model as a generator $G$. We use GAIL instead of other IL algorithms as it can best approximate the expert policy by minimizing a real metric defined on policy space and is more efficient compared to other IL algorithms \cite{ho2016generative}. The algorithm suited for our task is briefed in Algorithm 1. Compared to the original GAIL, we change the discriminator loss function to least square objective \cite{mao2017least} for better performance. Therefore, we denote our framework as Least Square-GAIL or LSGAIL.

\begin{algorithm}[t]
\caption{LSGAIL for conversational search policy learning}\label{alg:cap}
\begin{algorithmic}
\State \textbf{Input} Expert trajectories $\tau_E\sim\pi_E$, initialized policy $G_{\theta_0}$, initialized discriminator $D_{\sigma_0}$
\State \textbf{for} $i = 0,1,2,...$ \textbf{do}
\State \quad    Sample trajectory $\tau_i\sim\pi_{\theta_i}$
\State \quad    Update $\sigma_i$ by minimizing target:
\State \quad \quad $\sum_{\tau_i}D_{\sigma_i}(s,a)^2+ \sum_{\tau_E}(D_{\sigma_i}(s,a) - 1)^2$
%$E_{\tau_i}[\log( D_{\sigma_i}(s,a))] + E_{\tau_E}[\log (1-D_{\sigma_i}(s,a))]$
\State \quad    Update $\theta_i$ by maximizing KL-constrained target:
\State \quad \quad  $\text{E}_{\tau_i}[\log (G_{\theta_i} (a|s) * Q(s,a))] -\lambda H(\pi_{\theta_i})$,
\State \quad \quad where $Q(s,a) = E_{\tau_i}[\log (D_{\sigma_i}(s,a))|s_0=s, a_0=a]$
\State \textbf{End for}
\end{algorithmic}
\end{algorithm}

GAIL algorithm requires expert demonstration $\tau_E$ as input, which is a trajectory of \texttt{(state, action)} pairs that can provide the best conversational search quality. $\tau_E$ can be represented as $\tau_E=\{(s_0,a_0), .., (s_T, a_T)\}$. However, most conversational search datasets only have the raw conversations which can be represented as $\{U_0, A_0,..,U_N,A_N\}$. To get $\tau_E$, we evaluate all possible conversational search trajectories in terms of resulting conversational search quality and select the best as the expert demonstration. The metric we use to evaluate the conversational search quality will be discussed in \S\ref{sec:experiment}. The sampled trajectory $\tau_i$ is the trajectory decided by the policy model parameters in the current iteration.

The input of the \texttt{(state, action)} discriminator $D$ in the GAIL algorithm is the concatenated vector $S_a$ of the initial query $iq$, current utterance history $q$, the candidate text of the action, and the candidate ranking scores $s$ of the action. For example if the action is to return the result, then $S_a = (\textit{iq},q,\textit{re}_1,...\textit{re}_K, s_{\textit{re}}^{1:K})$. If the action is to ask the clarifying question, then $S_a = (\textit{iq},q,\textit{cq}_1,...\textit{cq}_K, s_{\textit{cq}}^{1:K})$. The output of $D$ is a scalar indicating the probability of the \texttt{(state, action)} pair being from the expert trajectory. $D$ is a 2-layer feedforward neural net:

\begin{equation}
D(S_a) = \text{Sigmoid}(W_{D2}\cdot \phi(W_{D1}\cdot S_a + b_{D1}) + b_{D2})
\end{equation}where $W_{D1}, W_{D2}, b_{D1}, b_{D2}$ are learnable parameters.

Our policy module is theoretically superior to previous RL-based conversational search models \cite{wang2021controlling, wang2022simulating} as the training of discriminator $D$ and policy $G$ is reward-free, thus avoiding the burdensome reward designing and tuning steps. The idea behind GAIL is to train $D$ with expert demonstrations and demonstrations generated by $G$, and use $\log(D(s, a))$ as the estimated reward during the update of $G$. %using the TRPO \cite{schulman2015trust} rule. Trust region policy optimization (TRPO) is an effective optimization method for large non-linear policy models. TRPO is different from line search optimization methods such as gradient descent. Instead of taking a step towards the gradient direction, it first determines a trust region with a max step size and then updates policy model to the optimum within the trust region. TRPO tends to give monotonic improvement for the policy model and prevent performance collapse during policy learning. 

\subsection{Inference}
\label{sec:inference}
During inference, an initial user query is first fed into the conversational search system; the system then enters the interaction loop with the user, where it will decide and make an action between asking clarifying questions and returning the search result to the user. The result and question retrieval models are run once to generate the input for the policy model, and then the policy model will make the final decision. The interaction loop will continue if the user responds to the clarifying question, and will stop if the system returns the search result or its retrieved clarifying question list does not contain any useful questions. Finally, the output of our system is a simulated conversational search trajectory which is a series of clarifying question and search result retrieval results and can be represented as $\{\pi_{\textit{cq}}^{1:T-1}, \pi_{\textit{res}}^T\}$, where $\pi_{\textit{cq}}^{1:T-1}$ represents the retrieved clarifying question ranklists for the first $(T-1)$ clarifying turns, and $\pi_{\textit{res}}^{T}$ is the retrieved result ranklist for the last turn. We will evaluate our system using the entire trajectory it produces. 

\section{Evaluation Metrics}
\label{sec:metric}
\subsection{\textbf{Expected Conversational Reciprocal Rank}}
In \S\ref{sec:algorithm}, we briefly mentioned an evaluation metric we use to select the expert demonstration from all possible conversation trajectories. We will explain this metric which we use for both selecting the expert demonstration and evaluating the performance of our system and the baseline models during inference.

The output of our conversational search system is a conversational search trajectory which can be represented as $\{\pi_{\textit{cq}}^{1:T-1}, \pi_{\textit{res}}^T\}$, where $\pi_{\textit{cq}}^{1:T-1}$ represents the retrieved clarifying question ranklists for the first $(T-1)$ clarifying turns, and $\pi_{\textit{res}}^{T}$ is the retrieved result ranklist for the last turn. With the above notations, we define the evaluation metric Mean Expected Conversational Reciprocal Rank based on Cascade model (Mean ECRR) as:
    \[\text{Mean ECRR} = \frac{1}{M}\sum_{i=1}^{M} \text{ECRR}_{\mathcal{U}}(\pi_{\textit{cq}}^{1:T-1}, \pi_{\textit{res}}^T)\] 
where ECRR is the Expected Conversational Reciprocal Rank for single conversation, and $\mathcal{U}$ is a cascade model to simulate user behavior in response to a retrieved clarifying question list. This is very similar to previous works such as \cite{chapelle2009err}, which also uses a user cascade model in their metric. To better understand cascade model and ECRR, we recommend imagining the term \textit{user} not as a single user but the entire user base instead. The cascade model assumes that (1) user will examine the clarifying question list by its rank order, and have a probability $\alpha$ to examine each clarifying question; (2) when seeing a relevant clarifying question, user will answer the question and wait for next response from the conversational system; (3) when seeing an irrelevant answer or clarifying question, user will continue to examine the next response with probability $\alpha$ or leave the conversation with probability $(1-\alpha)$. The parameter $\alpha$ reflects the percentage of users we assume to be tolerant and patient in the user base.

With the cascade model assumption, during all but the last conversation turn $t\leq (T-1)$, when a retrieved clarifying question ranklist $\pi_{cq}^t$ is returned to user, we can compute the following probability 

\begin{equation}
    p^t = \alpha^{r(\pi_{cq}^t=1)}
\end{equation} 
This is the probability that user will examine the true relevant clarifying question using cascade model on $\pi_{cq}^t$, where $r(\pi_{cq}^t=1)$ denotes the rank of relevant clarifying question. Then, user will either examine the clarifying question and continue the conversation with this probability $p^t$, which will leave the quality of the conversation to be further determined by future turns, or leave the conversation with probability $(1-p^t)$ without seeing a result, which is a disappointing scenario for both the user and our system. The above process can be described using the following equation:
    \begin{equation}
        \text{ECRR}^t = p^t\cdot\text{ECRR}^{t+1} + (1-p^t)\cdot0
    \end{equation}
        
In the last turn, when a retrieved result ranklist $\pi_{\textit{res}}^T$ is returned to user, we can finally evaluate the retrieved result ranklist quality using reciprocal rank of the true relevant result:
    \begin{equation}
        \text{ECRR}^T = \text{RR}(\pi_{\textit{res}}^T)
    \end{equation}
From equation (1) and (2), we can derive that the ECRR of the entire multi-turn conversation is computed as:
    
    \begin{equation}\label{ecrr definition}
        \text{ECRR}_{\textit{cas}}(\pi_{\textit{cq}}^{1:T-1}, \pi_{\textit{res}}^T) = \prod_{t=0}^{T-1} \alpha^{r(\pi_{\textit{cq}}^t=1)} \cdot \text{RR}(\pi_{\textit{res}}^T)
    \end{equation}
As mentioned earlier, the Ubuntu Dialog Corpus dataset only contains raw conversations. To train the imitation learning framework, we need to generate and compute expert trajectories from the raw data using ECRR. This can be done in the following steps (using one conversation as an example):
\begin{enumerate}
    \item In each turn of the conversation, we run both of the retrieval models to generate a result rank list $\pi_{\textit{res}}$ and clarifying question rank list $\pi_{\textit{cq}}$. 
    \item Assume the conversation has $T$ turns, then there are $T$ possible trajectories $\{\{\pi_{\textit{cq}}^{1:t-1}, \pi_{\textit{res}}^t\}^{t=1:T}\}$ in total, each ends at one of the $T$ turns. We use formula (9) to compute the ECRR of all these trajectories, and then select the trajectory with the highest ECRR as the expert trajectory. 
    
    Notice that different $\alpha$ can lead to different expert trajectories. For example, if we have a 2-turn conversation, and the retrieved results reciprocal rank for the 1st and 2nd turn are 0.33, 1, and the retrieved question rank for the 1st turn is 3. Using ECRR with $\alpha=0.5$, the expert trajectory is to not ask the clarifying question, because asking the question has ECRR = $0.5^{3} \times 1=0.125 < 0.33$. Using ECRR with $\alpha=0.7$, the expert trajectory is to ask the clarifying question, since $0.7^{3}\times 1 = 0.343 > 0.33$.
\end{enumerate}
ECRR can be seen as the average user satisfaction given the conversational search system response trajectory. This metric evaluates the full trajectory without assuming any specific user type. The entire user population is modeled by the parameterized cascade model. Some users may go through the full trajectory and receive a search result and other may not. Because of this, this metric is more realistic and applicable compared to the metrics used in \cite{wang2022simulating}.

According to the definition of ECRR (Equation ~\ref{ecrr definition}), the final score is determined by the number and quality of clarifying questions (cumulative product term) and the final retrieval result quality (reciprocal rank term). A policy trained to optimize ECRR is encouraged to get the best retrieved result (high effectiveness) while minimizing the number of clarifying questions asked to user (high efficiency). Thus ECRR accounts for both search effectiveness and efficiency. The parameter $\alpha$ in the cascade model of ECRR controls the trade-off between effectiveness and efficiency.

As discussed in section~\ref{sec:related}, most existing works about conversational search system use single-turn metrics or stack single-turn metrics under naive conversation policy, our work is among the few works which evaluate conversational search with multi-turn metrics. Among these few works, the majority of them uses multi-turn evaluation metrics only for evaluation, while we propose to also use the evaluation metric during conversational search policy training.

\subsection{\textbf{Recall and MRR}}

Besides the ECRR metric, we also include recall and MRR as other evaluation metrics, which are more commonly used when evaluating ranked lists. However, it is important to notice that we use them on the entire conversational search system trajectory $\{\pi_{cq}^{1:T-1}, \pi_{res}^T\}$, instead of any single turn ranking. In this case, it is also worth mentioning that MRR is a special case of ECRR when the cascade user model has \textbf{binary} $\alpha$. Specifically, $\alpha=1$ when the top question is relevant, and $\alpha=0$ otherwise. which means that users will always leave on seeing irrelevant clarifying questions.

\section{Experiments and Evaluations}
\label{sec:experiment}
\subsection{Experiment Design}
The goal of our experiments is to show that the proposed reward-free IL framework can (1) work reasonably well without reward tuning, and (2) generalize to different evaluation metrics or user assumptions. To show (1), we compare our framework with a risk-aware conversational search policy model Risk-aware Conversational Search agent with Q-learning (RCSQ) \cite{wang2022simulating}, which needs manually tuning the reward. We will denote this model as RCSQ for the rest of our paper. With the retrieval models fixed, we run our framework and the RCSQ model with different rewards and compare their performances using various evaluation metrics. To show (2), we train our conversational search policy regarding different evaluation metrics and compare its performance with the RCSQ model performance on these different evaluation metrics.

\subsection{Dataset}

% may need to change this to UDC
We use the Ubuntu Dialog Corpus (UDC) dataset \cite{udcdataset} in our experiments. The UDC dataset consists of question answering conversations about the Ubuntu system. We adopt the same processed and filtered dataset and simulation experiments as \cite{wang2021controlling, wang2022simulating}. The processing and filtering of the UDC dataset ensure that (1) all the conversations involve only two participants, i.e., the user and Microsoft agents, and in turns, (2) all the conversations have at least two turns, which means that there will be at least one clarifying question in each conversation. We use the filtered and randomly sampled 10000 conversations from \cite{wang2022simulating} as our dataset. Further details of the dataset can be found in Table~\ref{datasetstat}.

\begin{table}[t]
\caption{Original Ubuntu Dialog and our dataset statistics}
\vspace{-10pt}
\begin{center}
\resizebox{0.85\columnwidth}{!}{
\begin{tabular}{l|l|l}
\toprule
item             & Ubuntu Dialog & Ours  \\
\hline
\#conversations & 930,000 & 10,000  \\
\hline
Max. turns        & - & 10     \\
\hline
Min. turns        & 3 & 4       \\
\hline
Avg. turns        & 7.71 & 4.77  \\
\hline
Avg. \#words per utterance  & 10.34 & 20.85  \\
\bottomrule
\end{tabular}
}
\label{datasetstat}
\end{center}
\end{table}

\subsection{Baselines}
Following our experiment design, we include two groups of baseline models in our experiments. The first group comprises 3 naive conversational search policies and 1 simple policy learning model:

\begin{enumerate}
\item \textbf{Q0A}, a baseline that always returns results without asking a clarifying question. Hence, it always uses the initial query as the only information for answer retrieval. 
\item \textbf{Q1A}, a baseline that always returns results after asking exactly one clarifying question. It will always ask a clarifying question in the 1st turn, and then return results. 
\item \textbf{Q2A}, a baseline similar to Q1A, but it will always return results in the 3rd turn after asking clarifying questions in the first two turns. 
\item \textbf{CtxPred}, a simple conversational search policy learning model using behavior cloning \cite{behaviorcloning}. This is similar to the clarifying-question-need classification models in works such as \cite{xu2019asking, aliannejadi2020convai3}.

\end{enumerate}

The second group of baselines comprises a few untuned variations of the RCSQ model in \cite{wang2022simulating}. We use RCSQ as a representative for reinforcement learning policy models in general. The meaning of having untuned versions of RCSQ is for the comparison between the proposed reward-free IL framework and reward-tuning methods such as RCSQ.

Finally, we include an oracle policy that can foresee all possible conversation trajectories and can always take the best action by backtracking. It is worth mentioning that a conversational search policy does not improve the retrieval models it works with. It only decides which retrieval model output (results or questions) to return to the user. Hence, given fixed retrieval models, there will be an upper bound for any conversational search policy. This oracle policy is the upper bound and is only meant to be used for reference. 

\subsection{Technical Details}
We use the poly-encoder implementation from ParlAI\footnote{https://github.com/facebookresearch/ParlAI/tree/master/projects/polyencoder}. We download their pretrained checkpoints and finetune them on the UDC dataset. We implement our proposed IL framework from scratch based on Pytorch. We use the RCSQ model implementation from their GitHub repository. %\footnote{https://github.com/zhenduow/conversationalQA}. 
Our main experiments are run on a single-core GeForce RTX 2080 Ti with 11GB of memory. The pre-training of the poly-encoder retrievers is done on 4 of the above cores with batch size = 8.

\begin{table}[t]
\caption{RCSQ model reward table}
\vspace{-10pt}
\begin{center}
\resizebox{0.8\columnwidth}{!}{
\begin{tabular}{c|c|c}
\toprule
& Relevant & Irrelevant \\\hline
Search results & \multicolumn{2}{c}{Result Reciprocal Rank} \\
\hline
Clarifying question & $r_{cq}=0.11$ & $p_{cq} = -0.89$ \\
\bottomrule
\end{tabular}
}
\label{RCSQ reward table}
\end{center}
\vspace{-5pt}
\end{table}

In our experiments, we use the split dataset of train/val/test sets using 8:1:1 ratio in \cite{wang2022simulating}. The number of negative samples $k=99$. The original RCSQ model uses $r=0.11$ as their tuned reward. We test their model with untuned rewards $r=-0.1, 0.3, 0.5, 0.7, 0.9$. We train our model with a learning rate $lr=10^{-4}$, and entropy weight $\lambda=10^{-2}$. In each epoch of GAIL training, we will train the discriminator 5 times and the generator 1 time.

\section{Results and Analysis}
\label{sec:result}

% explaining table layout
\subsection{Experiment Results}
Our experiment results are shown in Table~\ref{udcpoly}. The rows are different conversational search policies. The 1st to 4th rows are the naive baseline models. Q0A, Q1A, and Q2A are determined policies, which ask exactly 0, 1, and 2 questions respectively before returning results to the user. CtxPred is a behavior cloning baseline that is trained using expert trajectories. These 4 baseline models in the first block represent the performances when using the state-of-the-art dense retrieval models directly without a conversational search policy. The second block is the untuned risk-control policies \cite{wang2022simulating} denoted as RCSQ. For example, the row of RCSQ $r=0.11$ is RCSQ with a reward table shown in Table~\ref{RCSQ reward table}. The third block is our proposed IL framework trained regarding different training targets, denoted as LSGAIL. For example, the row of LSGAIL $\alpha=0.3$ is LSGAIL trained under ECRR with cascade $\alpha=0.3$, where $\alpha$ is the frequency that the user will remain in the conversation after seeing an irrelevant clarifying question. The last row is the oracle policy, which is the performance upper bound of any conversational search policy, given the fixed retrieval models. This performance upper bound is only meant to be used for reference, since it is almost unachievable.
\begin{table*}[t]
\caption{Comparison of all models and baselines on sampled UDC dataset using poly-encoder as re-ranker. GAIL model is trained using ECRR cascade user model $\alpha=0.3,0.5,0.7, 0.9$. $\rho$ is the user patience for total clarifying questions, $\tau$ is the user tolerance for irrelevant clarifying questions. Numbers in bold mean the result is the best among all the variation. $\dag$ and $\ddag$ means $p < 0.1$ and $p < 0.05$ statistical significance over the all other models. $\star$ denotes the model variation used by \cite{wang2022simulating}.}
\vspace{-5pt}

\begin{tabular}{c|c|c|c|c|c|c}
\toprule
Policies  & R@1/100 (binary $\alpha$) & MRR (binary $\alpha$)  & ECRR $\alpha$ = 0.3  & ECRR $\alpha$ = 0.5 & ECRR $\alpha$ = 0.7 & ECRR $\alpha$ = 0.9\\ \hline  
Q0A     & \textbf{0.1580}   & \textbf{0.2381} & \textbf{0.2381}  & \textbf{0.2381} & \textbf{0.2381} & 0.2381\\ 
Q1A     & 0.1200   & 0.1630 & 0.1799 & 0.1938 & 0.2214 & \textbf{0.2693} \\ 
Q2A     & 0.0230   & 0.0284 & 0.0359 & 0.0387 & 0.0530 & 0.0761\\ 
CtxPred & 0.1165   & 0.1581 & 0.1531  & 0.1841 & 0.1951 & 0.2182\\ \hline
RCSQ r=-0.1& 0.1580 & 0.2381 & 0.2381 & 0.2381 & 0.2381 & 0.2381\\
RCSQ r=0.11$^\star$ & 0.1675 & 0.2449 &  0.2452  & 0.2479 & 0.2524 & 0.2612 \\  %epoch 4
RCSQ r=0.3 & 0.1700 & \textbf{0.2504}$^{\ddag}$  & \textbf{0.2471}$^{\ddag}$ & \textbf{0.2495}$^{\ddag}$ & \textbf{0.2535}$^{\ddag}$ &  0.2613\\ %epoch 7
RCSQ r=0.5 & \textbf{0.1710}$^{\ddag}$ & 0.2348 & 0.2190 & 0.2291 & 0.2455 & \textbf{ 0.2767}$^{\ddag}$ \\ %epoch 7
RCSQ r=0.7 & 0.1630 & 0.2204 & 0.2015 & 0.2262 & 0.2347 &  0.2735 \\ %epoch 5
RCSQ r=0.9 & 0.1470 & 0.1999 & 0.1771 & 0.2220 & 0.2263 & 0.2764 \\\hline
LSGAIL $\alpha=0.3$ & 0.1600  & 0.2404 & \textbf{0.2397}  & 0.2399 & 0.2403 & 0.2407 \\ %epoch 43
LSGAIL $\alpha=0.5$ & \textbf{0.1630} & \textbf{0.2410} & 0.2346  & \textbf{0.2405} & 0.2408 & 0.2412\\  %epoch 57
LSGAIL $\alpha=0.7$ & 0.1620 & 0.2406 & 0.2394 & 0.2400 & \textbf{0.2410} & 0.2428\\ % epoch 70
LSGAIL $\alpha=0.9$ & 0.1600 & 0.2123 & 0.1929 & 0.2066 & 0.2286 & \textbf{0.2696}\\\hline %epoch 42
Oracle  & 0.2330    & 0.3233  & 0.3152  & 0.3208 & 0.3298 & 0.3570 \\ 
\bottomrule
\end{tabular}
\label{udcpoly}

\vspace{-5pt}
\end{table*}

\begin{figure*}[t]
    \vspace{-5pt}
    \centering
    \begin{subfigure}[b]{0.25\textwidth}
    \centering
    \includegraphics[width=\textwidth]{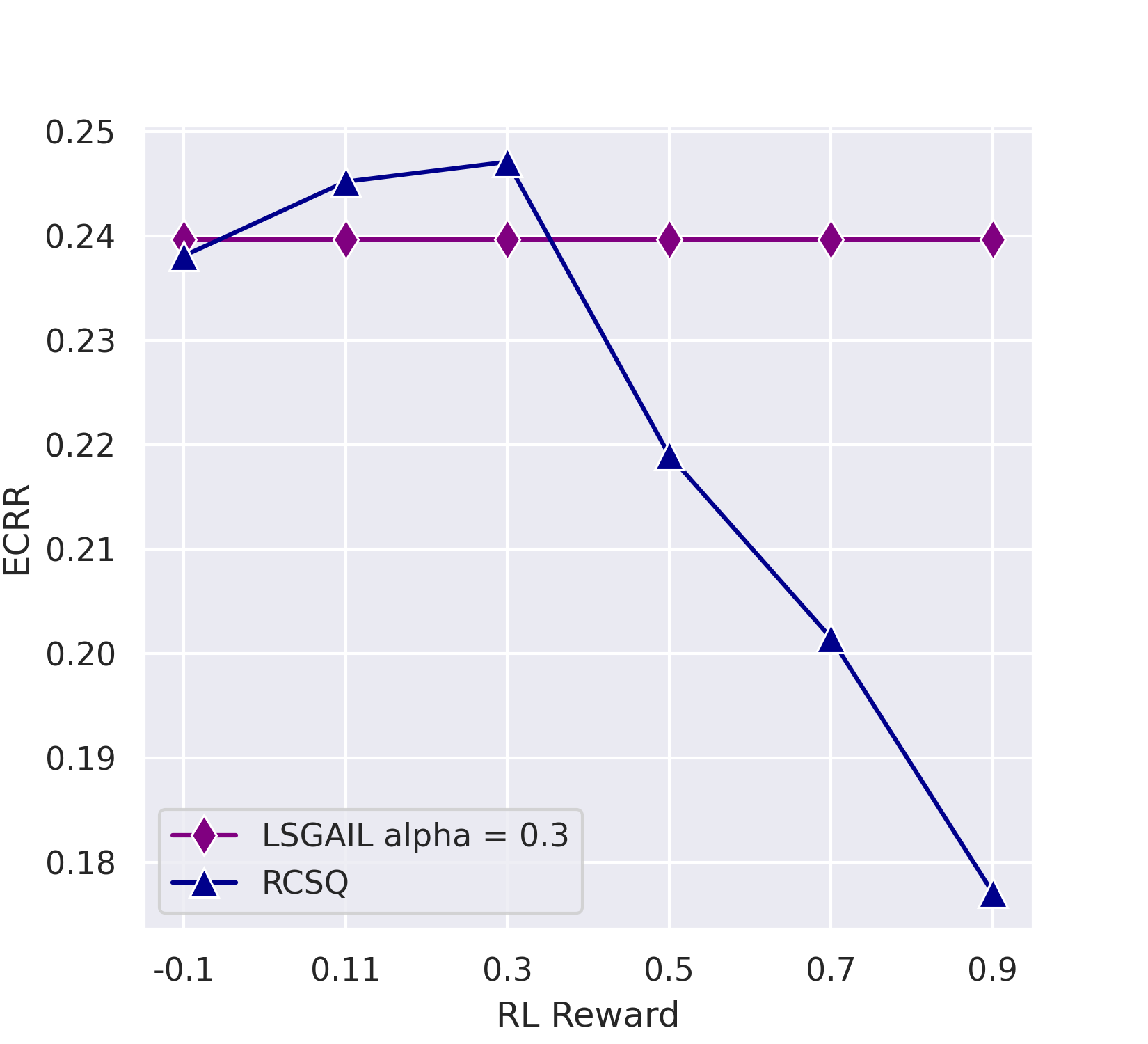}
    \caption{ECRR $\alpha=0.3$}
    \label{fig:0.3}
    \end{subfigure}
    \begin{subfigure}[b]{0.23\textwidth}
    \centering
    \includegraphics[width=\textwidth]{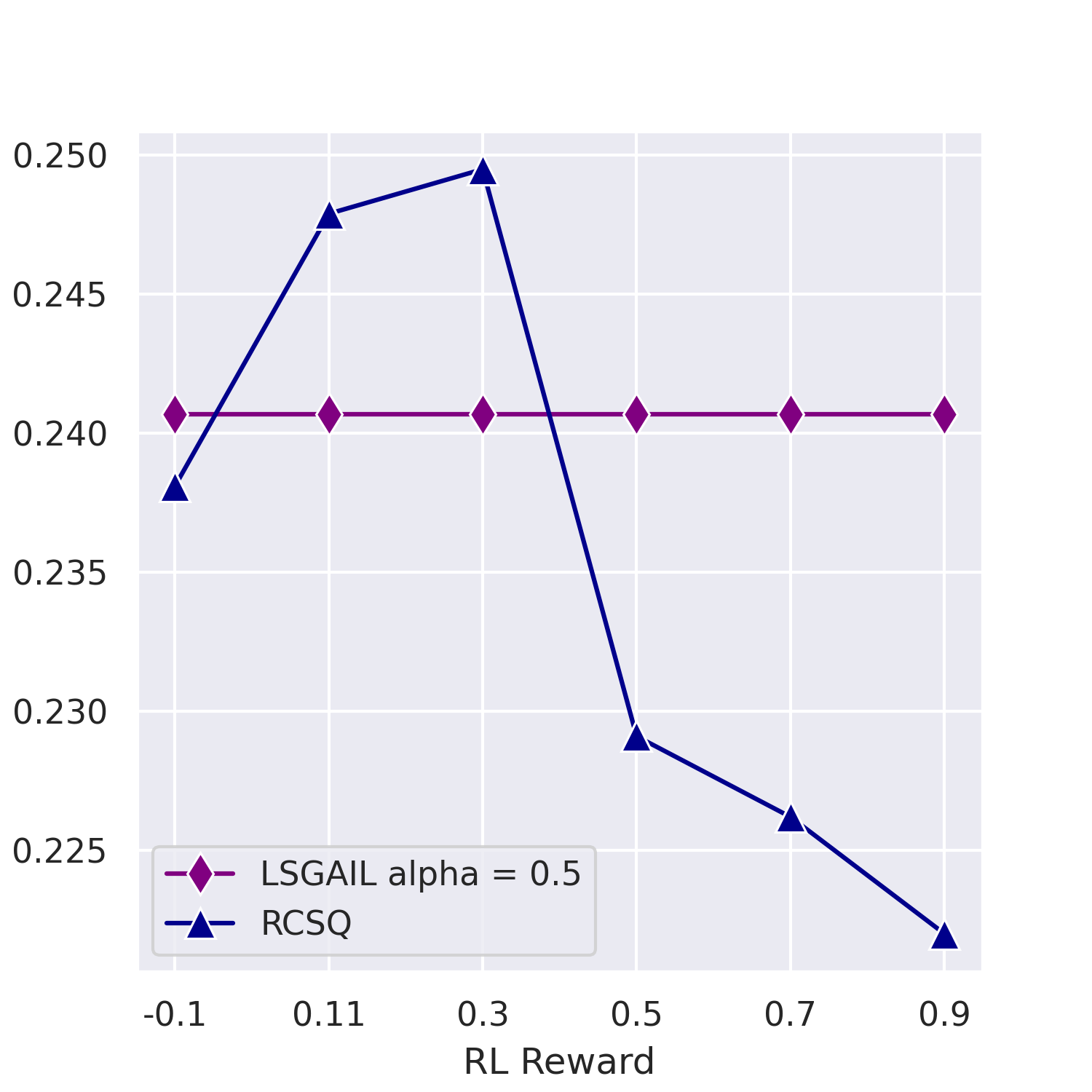}
    \caption{ECRR $\alpha=0.5$}
    \label{fig:0.5}
    \end{subfigure}
    \begin{subfigure}[b]{0.23\textwidth}
    \centering
    \includegraphics[width=\textwidth]{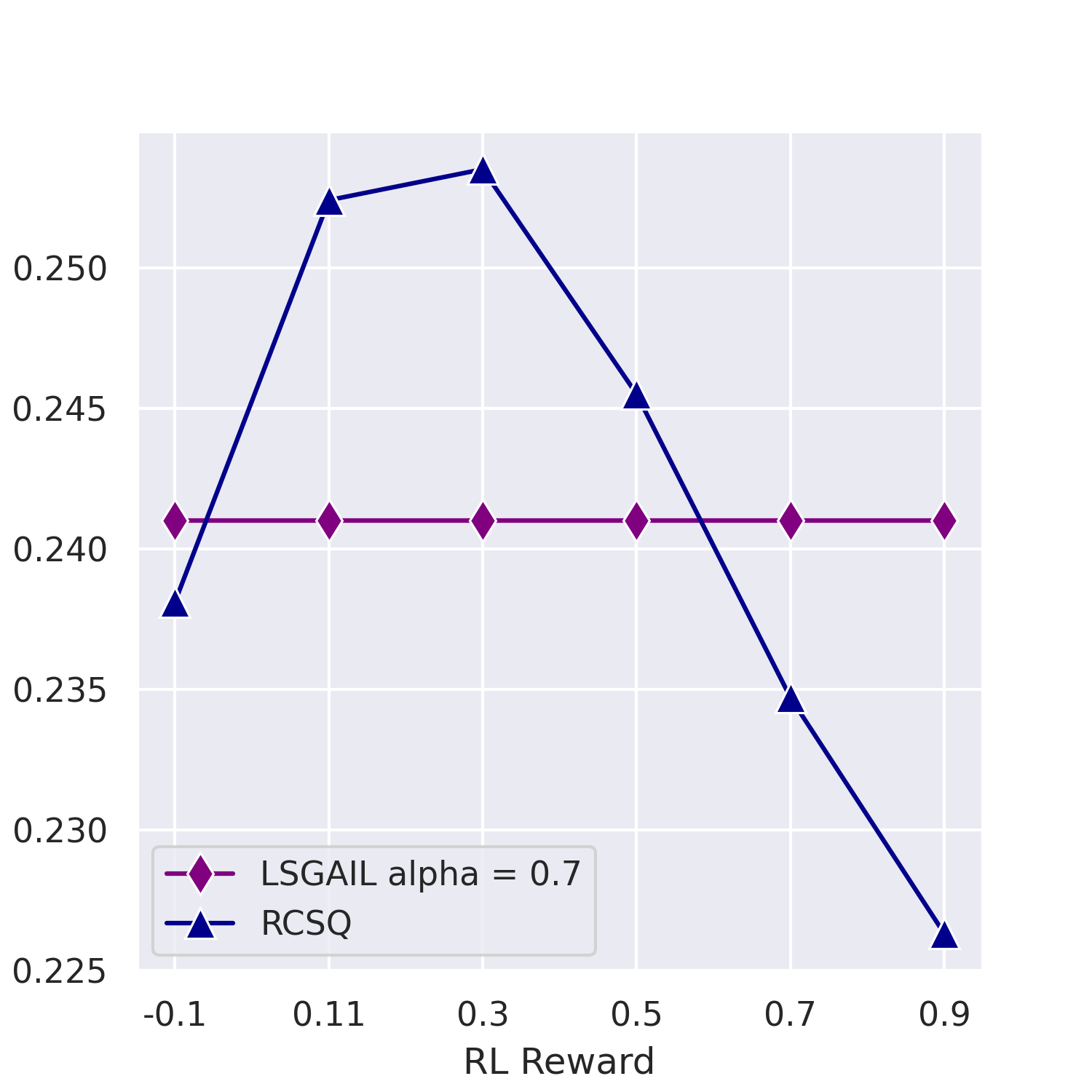}
    \caption{ECRR $\alpha=0.7$}
    \label{fig:0.7}
    \end{subfigure}
    \begin{subfigure}[b]{0.23\textwidth}
    \centering
    \includegraphics[width=\textwidth]{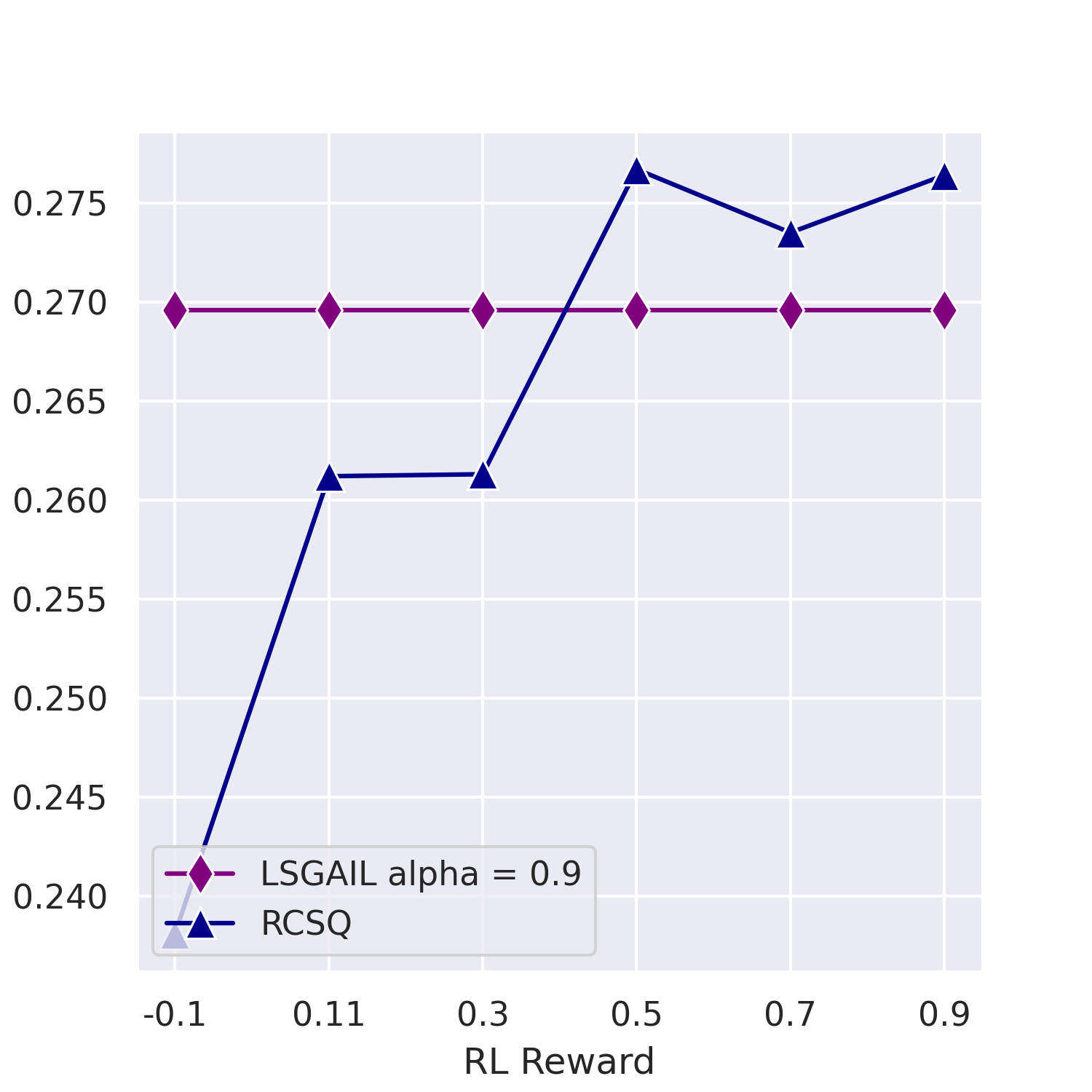}
    \caption{ECRR $\alpha=0.9$}
    \label{fig:0.9}
    \end{subfigure}
    \vspace{-5pt}
    \caption{Comparing LSGAIL with different $\alpha$ and RCSQ model with different rewards.}
    \label{fig:comparison}
    \vspace{-5pt}
\end{figure*}

The columns are different evaluation metrics. R@1/100 means recall@1 from 100 candidates, and MRR is the mean reciprocal rank. Notice that they are both computed on the entire conversational search trajectory. Followed by them are 4 different ECRRs with $\alpha=0.3, 0.5, 0.7, 0.9$. From the table, we can see that each LSGAIL variation performs the best on ECRR with the same $\alpha$ (bolded in the table). This shows that different LSGAIL variations are trained indirectly using different evaluation metrics as their training targets for conversation policy. Most of the time, LSGAIL can train the policy to optimize the designated evaluation metric.

Now, we answer three research questions to explain the result table and explain why the results can show our claims in the experiment design are positive. 

\subsubsection*{\textbf{RQ1: How does different user types ($\alpha$) affect RCSQ?}}
From Table~\ref{RCSQ reward table}, we can see that none of the RCSQ models give the best performing policy on all the ECRR metrics. The tuned RCSQ with $r=0.3$ outperforms other models on all the metrics except for Recall@1/100 and ECRR $\alpha=0.9$. The best performing model on these two metrics is RCSQ with $r=0.5$. We can also see how the performances of RCSQ with different rewards vary from Fig~\ref{fig:comparison}.

The fact that no single universal RCSQ reward can fit all of the user types is why RCSQ is not good enough for conversational search policy learning. We now explain why this result is reasonable. The idea behind the RCSQ model is to model conversational search policy as a risk-controlling task in conversational search actions. (1) The parameter $r$ in RCSQ represents the reward for clarifying questions. Using a small reward makes the policy more conservative about asking clarifying questions, and a higher reward makes the policy more optimistic about asking questions. 
(2) The parameter $\alpha$ in ECRR reflects the actual degree of conversation action risks in the form of users' patience for clarifying questions. Smaller $\alpha$ represents impatient users, who will be more likely to leave the conversation when seeing bad clarifying questions. Larger $\alpha$ represents patient users, who will spend more time on clarifying questions and interact more with the system. 

As a result of (1) and (2), RCSQ with smaller $r$ should theoretically work better on ECRR with smaller $\alpha$, and worse on ECRR with larger $\alpha$, and vice versa. which is shown in our experiments (this can be seen from the figure and the cells where RCSQ with $r=0.5, 0.7, 0.9$ underperform the baselines on ECRR $\alpha=0.3, 0.5$, while RCSQ with $r=0.11, 0.3$ underperform on ECRR $\alpha=0.9$). However, in real-world scenarios, we do not know ahead about the users of our search systems. Even if we do, no reward selection mechanism utilizes our knowledge about the users in RCSQ besides grid search. This makes RCSQ (and all reinforcement learning models) hard to generalize to various user scenarios.
\begin{table*}[ht]
\caption{An example of LSGAIL choosing the best search system actions.}

\begin{tabular}{p{0.02\linewidth}|p{0.02\linewidth}p{0.29\linewidth}|p{0.51\linewidth}}
\toprule
\multicolumn{1}{l|}{\textbf{Turn}} & \multicolumn{2}{c|}{\textbf{Conversation}} & \multicolumn{1}{c}{\textbf{Analysis}} \\ \hline
\multicolumn{1}{l|}{\multirow{3}{*}{Turn 1}} & \multicolumn{1}{l|}{User} & (Initial Query) Well i logged out and logged back in and somedude, the new user, still cannot sudo without being told off that this incident will be reported. & \multirow{3}{\linewidth}{This retrieved result is the correct result, but  is ranked 6th by the result retrieval model.
This clarifying question is relevant, and is ranked 1st by the question retrieval model.
In this case, the reward of returning the result is $1/6=0.167$, while the reward of asking the question is $r = 0.1<0.167$. Because of this, RCSQ chooses to directly return the result to the user.  Hence the ECRR of the RCSQ model is equal to $1/6 = 0.167$.  LSGAIL chooses to ask with the clarifying question ranked. Let's see what happens in the next turn.} \\ \cline{2-3}
\multicolumn{1}{c|}{} & \multicolumn{1}{l|}{Result}   &  /etc/sudoerrs is for handling SUDO... I don't know how yours is right now set, but check the content, maybe your default user was autmatically included. &                   \\ \cline{2-3}
\multicolumn{1}{c|}{} & \multicolumn{1}{l|}{Question} & Whom password do you use for sudoing? &          \\ 
\hline
\multicolumn{1}{l|}{\multirow{3}{*}{Turn 2}} & \multicolumn{1}{l|}{User} & (Reply) I use the password of the new user, somedude, that i created. & \multirow{3}{\linewidth}{The retrieved result is the same as in turn 1, but it is now ranked 1st by the result retrieval model thanks to more context. 
The retrieved question is also relevant, but it is ranked 7th by the question retrieval model. LSGAIL chooses to ask no more questions and returns the result, hence ECRR $=0.7^1*1=0.7$. If the system would ask the question, which is still relevant, and then go to turn 3, it would get an ECRR $=0.7^7*1\approx0.08$, which is lower because it would consume the user's time and degrade the user's search experience. } \\ \cline{2-3}
& \multicolumn{1}{l|}{Result}  & Same as in turn 1. &                   \\ \cline{2-3}
& \multicolumn{1}{l|}{Question} &  I see... I was asking because it's a common error using the root's password. Is the user listed in /etc/sudoerrs, or does it belong to the same group than your user?  &  \\
\bottomrule
\end{tabular}
\label{casestudy}
\end{table*}
\vspace{-10pt}

\subsubsection*{\textbf{RQ2: How does LSGAIL compare to RCSQ?}}
From Table~\ref{RCSQ reward table}, by comparing the performances of the LSGAIL framework and other policies vertically, we can draw the following conclusions: LSGAIL outperforms all the baseline policies most of the time. Using LSGAIL $\alpha=0.5$ as an example, LSGAIL outperforms all the baseline policies including Q0A, Q1A, Q2A, and CtxPred on R@1/100, MRR, and ECRR $\alpha=0.3$ significantly. From the table, the best baseline model is Q0A, with R@1/100 = 0.1580, MRR = 0.2381, and ECRR = 0.2381. LSGAIL gets R@1/100 = 0.1600, MRR = 0.2403, and ECRR = 0.2397, 0.2399, 0.2403. This implies that LSGAIL itself is an effective conversation policy training algorithm.

In general, LSGAIL performances are on par with RCSQ. When compared with RCSQ variations with rewards that are too small or too large, LSGAIL can achive better performances. This is shown in Fig~\ref{fig:comparison}. It does perform slightly worse than the finetuned RCSQ model variations with $r$=0.11 or 0.3 due to the lack of reward signal during training. However, when used on new dataset or unknown user types, LSGAIL does not need to manually finetune to find the best reward to achieve similar performances.

\subsubsection*{\textbf{RQ3: How does different user types ($\alpha$) change the comparison?}}
In RQ2, we see that LSGAIL can keep on par with RCSQ, although it always slightly underperforms the best RCSQ variation. In RQ3, we further study their performances by comparing the result table and Fig~\ref{fig:comparison} horizontally. Surprisingly, we cannot find any single RCSQ reward that always outperforms LSGAIL on all ECRR with different $\alpha$. Specifically, RCSQ with $r=0.1, 0.3$ are better than LSGAIL on ECRR with $\alpha=0.3, 0.5, 0.7$. However, they do not generalize to ECRR with $\alpha=0.9$, which represents the case with more patient users. To get good performance, RCSQ needs to finetune its reward to the range of $[0.5, 0.9]$. In contrast, LSGAIL never needs reward tuning and achieves comparable performances, regardless of small or large $\alpha$. This further shows that LSGAIL is theoretically more generalizable and easier to deploy than RCSQ.

\subsection{Case Study}

We conduct case studies to understand why LSGAIL can outperform the untuned RCSQ model. Table~\ref{casestudy} is an example we choose in the experiment that compares LSGAIL $\alpha=0.9$ and RCSQ $r=0.1$:

In this example, LSGAIL asks one clarifying question and gets ECRR = 1, RCSQ directly returns the result and gets ECRR = 0.167. From this example, we can see that RCSQ tends to only ask clarifying questions when the retrieved result reciprocal rank is lower than the finetuned reward. Because of this mechanism, RCSQ policy can improve the search result when it is completely irrelevant, e.g., when there are no relevant results in the top 10. However, its shortage is it cannot improve sub-optimal results to good results, e.g., improve the correct result from 6th to 1st. The training of LSGAIL does not set a hard reward for clarifying questions. Hence, it does not have the RCSQ policy problem, and it can further improve sub-optimal results to good results.

\section{Conclusions}
\label{sec:conclusion}
In this paper, we highlight the necessity of a conversational search policy in conversational search systems. As a solution, we propose a reward-free imitation learning framework for conversational search policy learning to address the problem in reinforcement learning methods. That is, they require heavy reward tuning, and are hard to generalize to different tasks and user types. The reward-free imitation learning framework trains the policy by inferring the best rewards from expert trajectories, which can be computed at a low cost from raw conversational search logs. The algorithm could potentially generalize to any user assumptions. Hence, it solves both of the two problems of reinforcement learning methods. 

To show our proposed framework can solve the two problems, we design experiments on the Ubuntu Dialog Corpus dataset and compare our proposed framework with three naive baseline policies, one behavior cloning policy learning method, and one representative reward-tuning reinforcement learning model. To evaluate the entire conversational search trajectory, we propose a new multi-turn evaluation metric called ECRR. Our experiment results show that our proposed framework can work reasonably well without reward tuning, and it can generalize well to different user assumptions. Our paper provides a useful reward-free imitation learning framework for conversational search policy training, which is easier to deploy than traditional reinforcement learning methods and more flexible to various user assumptions.


%\begin{acks}

% \section*{Acknowledgements}
% This work was supported in part by the School of Computing, University of Utah. Any opinions, findings and conclusions or recommendations expressed in this material are those of the authors and do not necessarily reflect those of the sponsor.

%\end{acks}


\bibliographystyle{ACM-Reference-Format}
\bibliography{sample-sigconf}

\appendix

\end{document}